\documentclass{PoS}
\usepackage{psfrag}

\newcommand{\be}{\begin{equation}}
\newcommand{\ee}{\end{equation}}
\newcommand{\bea}{\begin{eqnarray}}
\newcommand{\eea}{\end{eqnarray}}
\newcommand{\bi}{\begin{itemize}}
\newcommand{\ei}{\end{itemize}}

\newcommand{\Dov}{D_{\rm N}}
\newcommand{\pbf}{{\bf p}}

\title{Universality check of the overlap fermions in the Schr\"odinger functional}

\ShortTitle{Universality check of the overlap fermions in the Schr\"odinger functional}

\author{
\rightline{\rm HU-EP-08/29}
\rightline{\rm SFB/CPP-08-57}\vspace{-5mm}
        \speaker{Shinji Takeda}\\
        Humboldt Universitaet zu Berlin\\
        E-mail: \email{takeda@physik.hu-berlin.de}
}

\abstract{
I examine some properties of the overlap operator
in the Schroedinger functional formulated by L\"uscher
at perturbative level.
By investigating spectra of the free operator and
one-loop coefficient of the Schroedinger functional coupling,
I confirm the universality at tree and one-loop level.
Furthermore, I address
cutoff effects of the step scaling function and
it turns out that the lattice artifacts for the overlap operator
are comparable with those of the clover actions.
}

\FullConference{The XXVI International Symposium on Lattice Field Theory \\
		 July 14 - 19, 2008\\
		 Williamsburg, Virginia, USA}

\begin{document}

\section{Introduction}
As well known, the overlap fermions~\cite{Neuberger:1997fp} have
an exact chiral symmetry on the lattice,
that is, the fermion operator satisfies the Ginsberg-Wilson (GW) relation.
This is an ideal formulation to compute the quark condensate,
which is an important physical quantity as an order parameter
of the spontaneous symmetry breaking,
without running into the subtle additive renormalization problem.
A renormalized quark condensate
depends on the renormalization scale and scheme,
but, in order to avoid such dependences
it is usually convenient to choose a renormalization group invariant
(RGI) as a reference quantity,
\be
\Sigma_{\rm RGI}
=
\lim_{g_0\rightarrow 0}
{\cal Z}_{\rm P}(g_0)\Sigma_{\rm lat}(g_0),
\label{eqn:renormalization}
\ee
where I have used the fact that
thanks to the chiral symmetry on the lattice
the renormalization factors for the flavor singlet scalar
and the flavor non-singlet pseudo-scalar density are equivalent
even at a finite lattice spacing.
Recently, the JLQCD collaboration \cite{Fukaya:2007fb}
estimated the bare quark condensate,
and then performed the renormalization
through a non-perturbative scheme, so called, RI/MOM scheme.
However, in order to carry out the renormalization in a solid way,
here I will use more sophisticated scheme,
the Schr\"odinger functional (SF) scheme.
As well known this scheme can be defined non-perturbatively at massless point,
and can avoid the large scale problem.

Now let me explain how to carry out the non-perturbative
renormalization of the quark condensate
by making use of the SF scheme.
My final goal is the RGI condensate in eq.(\ref{eqn:renormalization}).
In order to get this quantity from the given bare quantity
the renormalization factor ${\cal Z}_{\rm P}(g_0)$ is required.
This factor can be obtained by
the following renormalization program which
is divided into three parts.
\be
{\cal Z}_{\rm P}(g_0)
=
\hat{Z}_{\rm P,SF}^{\rm PT}(\infty,\mu_{\rm PT})
U_{\rm P,SF}^{\rm NP}(\mu_{\rm PT},\mu_{\rm had})
Z_{\rm P,SF,ov}^{\rm NP}(g_0,a\mu_{\rm had}).
\ee
First, the left factor $\hat{Z}_{\rm P,SF}^{\rm PT}(\infty,\mu_{\rm PT})$
is required to remove the scale and scheme dependence of the quark condensate.
If the energy scale $\mu_{\rm PT}$ is so high then it is sufficient to
use perturbation theory to get this factor.
The middle factor represents the non-perturbative evolution
of $Z_{\rm P}$ from a low energy $\mu_{\rm had}$
to the high energy scale.
Actually a product of this factor and
the previous one was already calculated by
the ALPHA collaboration~\cite{DellaMorte:2005kg}
for $N_{\rm f}=2$ in the SF scheme.
Note that it is independent of the discretization,
that is, the lattice action.
The right factor, $Z_{\rm P,SF,ov}^{\rm NP}(g_0,a\mu_{\rm had})$,
is a renormalization
factor relating the bare quark condensate and a renormalized one
at a certain renormalization scale, which should be low energy
$\mu_{\rm had}$ in order to avoid large cutoff effects.
In fact, the last factor is a missing piece to get the RGI.
Since this factor depends not only on the scale and the scheme
but also the lattice action (now it is the overlap fermion),
first of all, one has to define the overlap fermion in the SF.

In the next section, I will briefly introduce
the formulation proposed by L\"uscher.
I will not compute $Z_{\rm P}$ in this report,
instead I will show some preparative studies,
spectra of free operator and universality check
at both tree and quantum level.
Furthermore I will address cutoff effects
for the overlap fermion at one-loop level.
More details about results shown in the report
can be found in Ref.~\cite{Takeda:2007ga}.

\section{Formulations of the overlap fermion in SF}

Since the boundary conditions of the SF are not
compatible with the chiral symmetry,
it is not so trivial to formulate the overlap fermions in the SF.
An important issue is how to break the GW relation
while keeping the boundary conditions (up to $O(a)$).
So far, three formulations have been proposed.
First one is an orbifolding construction by Taniguchi~\cite{Taniguchi:2004gf},
which contains some subtleties.
However, as Sint showed in Ref.~\cite{Sint:2005qz},
such subtleties can be partially removed by
introducing a chirally rotated version of the SF.
Nevertheless, this is rather technically involved.
In this report, I take rather simpler one,
so called universality formulation proposed by
L\"uscher~\cite{Luscher:2006df}.

Let me introduce the universality formulation briefly.
A massless operator is given by
\be
\bar{a}\Dov
=
1 - \frac{1}{2} ( U + \gamma_5 U^{\dag} \gamma_5),
\hspace{10mm}
\bar a=a/(1+s),
\ee
\be
U=A X^{-1/2},
\hspace{10mm}
X=A^{\dag}A + c a P,
\hspace{10mm}
A=1+s-aD_{\rm w},
\label{eqn:UA}
\ee
where $s$ is a tunable parameter to optimize computational costs,
and $D_{\rm w}$ is the Wilson operator in the SF.
A crucial difference of the overlap operator in the usual
lattice and that in the SF is the presence of an operator $P$
in the inverse square root.
This operator is supported near boundary and
plays an important role to produce the correct boundary conditions
in the continuum limit.
Due to the presence of $P$, the matrix $U$ is not unitary
anymore. Accordingly the overlap operator does not satisfy
the GW relation, instead it follows the modified relation,
\be
\gamma_5 \Dov + \Dov \gamma_5 =\bar{a}\Dov \gamma_5 \Dov+\Delta_B,
\ee
with a breaking term $\Delta_B$.
It is shown \cite{Luscher:2006df}
that this breaking term is exponentially suppressed
away from the boundary, therefore,
the chiral symmetry is approximately maintained in the bulk.
The coefficient $c$ in eq.(\ref{eqn:UA})
has an important role to cancel $O(a)$ corrections
of physical quantities, and it has a perturbative expansion
\be
c=c^{(0)}+g_0^2 c^{(1)}+O(g_0^4).
\label{eqn:ccc}
\ee
According to the original paper,
I set the tree value $c^{(0)}=1+s$, which is an optimal choice
for the tree level O($a$) improvement,
in the following calculations.

In the definition of the overlap operator,
there is the inverse square root.
Due to the presence of the background field,
the kernel of the inverse square root
is not  diagonal matrix anymore even in the free case
and even after  performing partial Fourier transformations.
Therefore I have to rely on the numerical approximation
even in the perturbative calculation.
To this end, I use the minimax polynomial approximation~\cite{Giusti:2002sm},
\be
X^{-1/2}_{\pbf}
\approx
\sum_{k=0}^{N}
c_k
T_k((2X_{\pbf}-v_{\pbf}-u_{\pbf})/(v_{\pbf}-u_{\pbf})),
\ee
where $T_k$ is the Chebyshev polynomial of degree $k$.
$X_{\pbf}$ is a kernel
in the time-momentum space whose size is $4(T/a-1)$ square
for a fixed spatial momentum configuration $\pbf$, and
the minimal and maximum eigenvalues of $X_{\pbf}$ are denoted by
$u_{\pbf}$ and $v_{\pbf}$ respectively.
The coefficient $c_k$ is determined
by the Remez algorithm to obtain the Minimax polynomial.
In the summation step, I use the Clenshaw sum scheme
in order to maintain numerical precisions.
An accuracy for the approximation is set to $10^{-13}$.
Given this accuracy, a ratio between the minimum and maximum eigenvalue
determine the degree of polynomial.
In my computation,
$u_{\pbf}/v_{\pbf}\sim 0.01$ for a typical case.
Then the degree of polynomial turns out to be $N\sim 100$.
This is just for a purpose to give some feeling.

\section{Spectrum of free operator}
\begin{figure}[t]
\begin{center}
  \begin{tabular}{cc}
  \psfragscanon
  \psfrag{g}[][][0.6]{GW ($L=\infty$)}
  \scalebox{.89}{\includegraphics{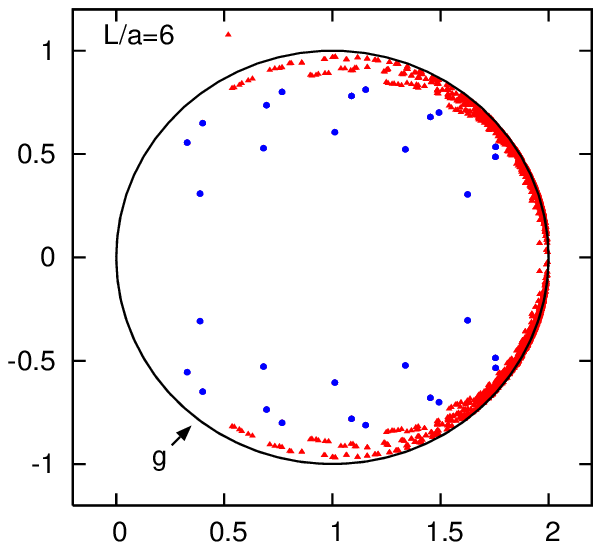}}&
  \hspace{-10mm}
  \scalebox{1.}{\includegraphics{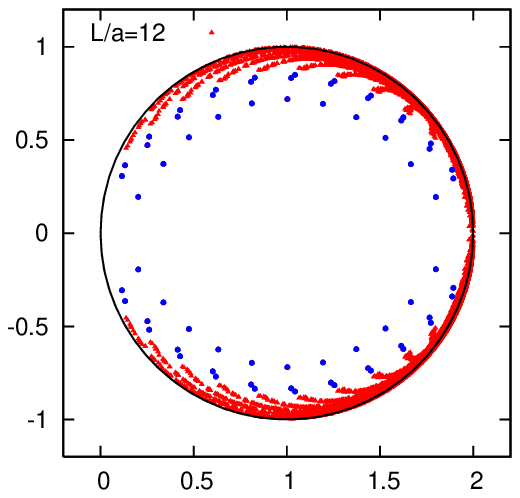}}\\
  \end{tabular}
\caption{Spectrum of $\bar{a}\Dov$ with $s=0$ and $\theta=0$
in the presence of the background field.
\label{fig:Dov}}
\end{center}
\end{figure}
Spectra of the free massless operator $\bar{a}\Dov$ are shown
in Figure~\ref{fig:Dov} for $L/a=6,12$.
The parameters are set to $s=\theta=0$ where
$\theta$ parameterizes the generalized boundary conditions
for the spatial directions.
The non-vanishing background gauge field \cite{Luscher:1992an}
is used here.
Blue points, which belong to a zero spacial momentum sector,
and red points, which are from the other sector,
represent individual eigenvalues.
Actually in the original paper~\cite{Luscher:2006df},
it is shown that the operator is bounded by a unit circle
\be
|| \bar{a}\Dov - 1 || \le 1,
\ee
and it is given by the black solid circle in the plot.
This equation indicates that all eigenvalues are contained in the circle.
On the other hand, in the infinite volume case, 
that is, the usual GW fermion, it is known that
all eigenvalues lie on the circle.
Therefore the deviation from the circle is considered
as boundary effects or finite size effects
and actually such deviation
is reduced for larger lattice in the right panel of Figure~\ref{fig:Dov}.
Especially I found that the blue points
are strongly affected by the boundary effects
in the sense that they are distant from the circle.

\begin{figure}[t]
\begin{center}
  \begin{tabular}{cc}
  \scalebox{1.1}{\includegraphics{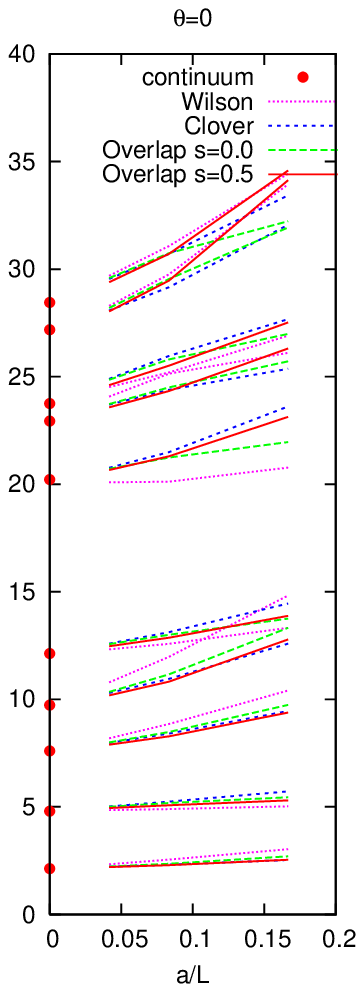}}&
  \scalebox{1.1}{\includegraphics{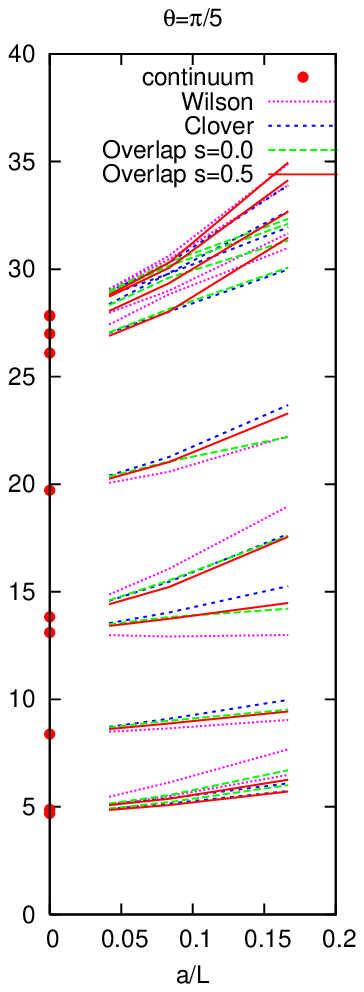}}\\
  \end{tabular}
\caption{Spectrum of $L^2\Dov^{\dag}\Dov$ with $\theta=0$ (left)
and $\theta=\pi/5$ (right)
in the presence of the background field.
The red points represent the continuum values from \cite{Sint:1995ch}.
\label{fig:Dov2}}
\end{center}
\end{figure}

Furthermore, I investigate spectra of hermitian operator $L^2\Dov^{\dag}\Dov$.
Actually the eigenvalues for this operator
have continuum limit~\cite{Sint:1995ch}.
Figure~\ref{fig:Dov2} shows scaling behaviors of the ten lowest eigenvalues.
All cases converge to the continuum limit properly for both $\theta$ values.
Therefore I conclude that the universality at the tree level is confirmed.

\section{Universality check in perturbation theory}
In the previous section I have investigated the property of the free operator.
In this section, let me address the universality at the quantum level.
For the purpose, I consider
the SF coupling~\cite{Luscher:1992an,Sint:1995ch}
\be
\bar{g}_{\rm SF}^2(L)
=
\left.
\frac{\partial \Gamma}{\partial \eta}
\right|_{\eta=\nu=0}
=
g_0^2[1+m_1(L/a)g_0^2+O(g_0^4)],
\ee
where $\Gamma$ is an effective action of the system
and the standard convention $T=L$ is taken.
Now I am interested in the one-loop contribution $m_1(L/a)$.
This is composed from the gauge and fermion parts,
$m_1(L/a)=m_{1,0}(L/a)+N_{\rm f}m_{1,1}(L/a)$,
and I compute the fermion part numerically
by using the overlap operator.
And then I analyze the data according to the Symanzik's expansion form
\be
m_{1,1}(L/a)
=
A_0+
B_0 \ln (L/a)+
A_1 a/L +
B_1 a/L \ln (L/a) +
O((a/L)^2).
\ee
I extract first few coefficients, $A_0$, $B_0$,...
by making use of the method in Ref.~\cite{Bode:1999sm}.

The first coefficient $A_0$ is generally a function of the parameter $s$,
and I get $A_0(s)|_{s=0}=0.012567(3)$,
while by combining the results of Ref.~\cite{Sint:1995ch,Alexandrou:1999wr},
$A_0(s)|_{s=0}=0.012566$
can be deduced. Consistency can be seen with a reasonable degree of accuracy.
$B_0$ is a coefficient of the log divergence
and this is related with the fermion part of the 
one-loop coefficient of the beta function,
$b_{0}^{\rm F}=-1/(24\pi^2)$.
I confirm $B_0=2b_{0}^{\rm F}$ up to 4 digits for several values of $s$.
From these results,
I can conclude that the universality at the quantum level is confirmed.
Furthermore, I determined $A_1$ as a function of $s$.
Actually this gives fermion part of the
O($a$) boundary counter term at one-loop order,
$c_{\rm t}^{(1)}=c_{\rm t}^{(1,0)}+N_{\rm f}c_{\rm t}^{(1,1)}$,
\be
c_{\rm t}^{(1,1)}
=
A_1/2
=
-0.00958-0.00206s-0.00484s^2-0.00748s^3-0.01730s^4.
\ee
This formula will be used for future simulations to
achieve one-loop O($a$) improvement.
I checked $B_1=0$ up to few digits.
Since this is a signal for tree level O($a$) improvement,
I can confirm this in an actual manner.

\section{Lattice artifacts of step scaling function}

\begin{figure}[t]
\begin{center}
  \scalebox{1.5}{\includegraphics{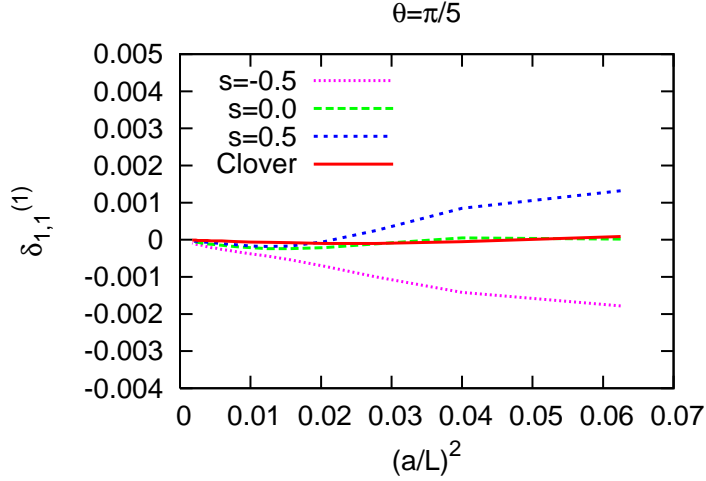}}
\caption{One-loop relative deviation as a function of $(a/L)^2$
with $\theta=\pi/5$.
\label{fig:deviation}}
\end{center}
\end{figure}

Finally let me show lattice artifacts of the step scaling function,
which describes the evolution of the running coupling,
\be
\sigma(u)=\bar{g}^2(2L),
\hspace{10mm}
u=\bar{g}^2(L).
\ee
The relative deviation is defined as
\be
\delta(u,a/L)
=
\frac{\Sigma(u,a/L)-\sigma(u)}{\sigma(u)}
=
\delta_1(a/L)
u + O(u^2),
\ee
where $\sigma(u)$ represents the step scaling function in the continuum limit
and $\Sigma(u,a/L)$ is that on the lattice, and
this tells us the size of lattice artifacts.
I evaluate the fermion part of this quantity to one-loop order,
$\delta_1(a/L)=\delta_{1,0}(a/L)+N_{\rm f}\delta_{1,1}(a/L)$.

Figure \ref{fig:deviation} shows its scaling behavior.
Note that the overlap fermion with $s=0$
shows almost flat, and this value of $s$
is an optimal choice from the point of view
of lattice artifacts.
For comparison, I include the results of
the clover action~\cite{Sint:1995ch},
and it also shows small cutoff effects.
Therefore I conclude that the lattice artifacts
of the clover and the overlap fermion with $s=0$ are comparable.

\section{Concluding remarks}
Among some formulations, I choose L\"uscher's formulation.
I investigate the spectra of the free overlap operators,
and then I observe the expected behaviors.
Next, I confirm the universality at quantum level,
and determine the O($a$) boundary counter term at
one-loop order, $c_{\rm t}^{(1,1)}$.
This is needed in future simulations to reduce cutoff effects.
Furthermore I investigate the lattice artifacts
of the step scaling function to one-loop order,
and then it turns out that the scaling behavior
of the overlap is comparable with the clover action.

As next targets, there are several quantities within
perturbation theory.
In this report, I exclusively consider the massless case,
however, I will investigate massive case too.
A comparison study with
the orbifolding formulations is also interesting.
Furthermore, still there is an improvement coefficient
which I have to compute
before starting simulations, $c^{(1)}$ in eq.(\ref{eqn:ccc}).
In fact the coefficients $c$
is only accessible within a framework of perturbation theory,
therefore two-loop calculations will be required.
In that course, it is very convenient to use the automatic method
developed last year~\cite{Takeda:2007dt}.

Finally, I have to remind readers that my final goal 
is the non-perturbative computation of $Z_{\rm P}$.

\vspace{7mm}

I thank the Deutsche Forschungsgemeinschaft (DFG)
for support in the framework of SFB Transregio 9.
I also thank FLAVIAnet for financial support.


\providecommand{\href}[2]{#2}\begingroup\raggedright\endgroup


\end{document}